\documentclass[rnote]{aa} 

\usepackage{savesym}
\savesymbol{iint}
\savesymbol{iiint}
\usepackage{graphicx}
\usepackage{txfonts}
\usepackage{natbib}
\usepackage{url}
\usepackage{amsmath}
\begin{document} 

   \title{ARES v2\thanks{Automatic Routine for line Equivalent widths in stellar 
                         Spectra - ARES Webpage: http://www.astro.up.pt/$\sim$sousasag/ares/} - new 
                         features and improved performance \thanks{Based on observations made with ESO Telescopes at the La Silla Paranal Observatory under programme ID 075.D-0800(A)}}
   \author{S. G. Sousa\inst{1}
          \and N. C. Santos\inst{1,}\inst{2}
          \and V. Adibekyan\inst{1}
          \and E. Delgado-Mena\inst{1}
          \and G. Israelian\inst{3,}\inst{4}
          }
          \institute{
          Instituto de Astrofísica e Ciências do Espaço, Universidade do Porto, CAUP, Rua das Estrelas, PT4150-762 Porto, Portugal
          \and Departamento de F\'isica e Astronomia, Fac. de Ci\^encias, Universidade do Porto, Rua do Campo Alegre, 4169-007 Porto, Portugal
          \and Instituto de Astrof\'isica de Canarias, 38200 La Laguna, Tenerife, Spain
          \and Departamento de Astrofisica, Universidade de La Laguna, E-38205 La Laguna, Tenerife, Spain
          }

   \date{Received September 15, 1996; accepted March 16, 1997}

 
  \abstract
  {}
   {We present a new upgraded version of ARES. The new version includes a series of interesting new features such as automatic 
   radial velocity correction, a fully automatic continuum determination, and an estimation of the 
   errors for the equivalent widths.}
   {The automatic correction of the radial velocity is achieved with a simple cross-correlation 
   function, and the automatic continuum determination, as well as the estimation of the errors, relies
   on a new approach to evaluating the spectral noise at the continuum level.}
   {ARES v2 is totally compatible with its predecessor. We show that the fully automatic 
   continuum determination is consistent with the previous methods applied for this task. 
   It also presents a significant improvement on its performance thanks to the implementation of 
   a parallel computation using the OpenMP library.}
  {}

   \keywords{Techniques: spectroscopic -- Methods: data analysis -- stars: solar-type -- stars: abundances}

   \maketitle
%

\section{Introduction}

One of the most commonly used spectral analysis methods is the line-by-line analysis, 
which in most of the cases requires a precise measurement of the strength of 
the spectral lines. This is normally done by measuring the equivalent-width (EW). 
EW methods have been used for decades in spectroscopy. 

In the past decade several automatic codes have emerged that aimed at releasing 
the astronomer from this tedious task (when using interactive routines), and, more importantly, 
to allow the precise and systematic measurement of equivalent
widths. ARES \citep[][]{Sousa-2007} was one of the first codes to appear freely available to the community. ARES measures the equivalent 
widths of lines in an automatic way, trying to follow as closely as possible the same steps as those that would be taken by an 
astronomer when using interactive routines. Several other codes exist and appeared more 
recently to perform the same task using different approaches: DAOSPEC \citep[][]{Stetson-2008}; DOOp, 
which is an automatic wrapper for DAOSPEC \citep[][]{CantatGaudin-2014}; TAME \citep[][]{Kang-2012}, which 
follows the basic ideas from ARES very closely, is written in IDL, and presents some interactive 
facilities for the spectral analysis. Rencently, a more complete spectral analysis 
code was presented that also computes EWs \citep[iSPEC - ][]{BlancoCuaresma-2014}.

Because it is so simple to use, ARES has been one of the most widely adopted codes for automatic 
EW measurements \citep[e.g.,][]{Hekker-2007, Neves-2009, Sozzetti-2009, Ghezzi-2010, 
Ruchti-2011, Sousa-2011, Adibekyan-2012, Tabernero-2012, 
Mortier-2013, Santos-2013, Monroe-2013, Maldonado-2013, Jofre-2014}.
ARES has also been used for asteroseismology studies that require stellar 
characterization \citep[e.g.][]{Creevey-2012, SilvaAguirre-2013, Metcalfe-2010} and is 
used in large spectroscopic surveys such as the GAIA-ESO Survey \citep[][]{Gilmore-2012} to analyze the high-resolution spectral data in several working groups of
the consortium (e.g., \citet[][]{Sousa-2014, Smiljanic-2014, Lanzafame-2015}). The 
algorithms in ARES are also being implemented in recipes for the very first 
data analysis software that will be available for a VLT instrument, 
ESPRESSO \citep[][]{Megevand-2012, DiMarcantonio-2012, Pepe-2014}.

We present the new improved version of the ARES code with special 
focus on the description of the new features that were introduced. \textbf{A more detailed description 
of new features is available on the ARES webpage.}


\section{New features}

\textbf{Correction for radial velocity}: In the previous version, the input spectra needed to 
    be fully corrected for radial velocity. Otherwise, ARES would not be able to correctly identify 
    the spectral lines that were to be measured. In the new version, the user has two options that can be 
    chosen to correct for the radial velocity in ARES: The first is to provide the value for 
    the radial velocity of the star in km/s. The second is to provide a mask with a list of 
    lines at rest frame that will then be used by ARES to estimate the radial velocity. To provide this 
    new input to ARES, the new parameter \textit{rvmask} needs to be included at the 
    very end of the input file \textit{mine.opt}. An example and a full description of the 
    method used for this new feature is provided on the ARES support 
    web-page\footnote{http://www.astro.up.pt/~sousasag/ares/support.html}.

\vspace{0.2cm}  
\textbf{Automatic \textit{rejt} estimation}: The parameter \textit{rejt} is the most 
    important input parameter for the proper normalization and is crucial for the correct EW 
    measurement. \textit{rejt} strongly depends on the signal-to-noise ratio (S/N). Originally, this 
    was a free parameter set by the user because the 
    visualization of the continuum position is highly subjective. This parameter continues to be free in ARES v2, 
    and we still recommend this as the best option for the user who desires to derive precise 
    and accurate EWs. In some of our previous works we have presented some suggestions for the best value for this parameter. First, in \citet[][]{Sousa-2008} we 
    provided a discrete relation between the spectrum S/N and the recommended \textit{rejt} value. 
    This table gives only values for S/N $>$ 100. \citet[][]{Sousa-2011} later provided 
    an empirical relation for the cases where the S/N is low (S/N $<$ 100). These relations 
    were derived for HARPS spectra, but according to our experience, they can be safely applied to other 
    spectra whose S/N was computed in a similar way. If the S/N is not 
    available, we recommend either adjusting this parameter individually for each 
    spectra or finding a way to first estimate the S/N for the spectra and then derive the 
    proper relation with the \textit{rejt} parameter \citep[for an example of such an
    exercise see][]{Mortier-2013}.

    The new version of ARES has an additional option where the user can specify spectral 
    regions in the spectra, which should ideally be clear of spectral lines. ARES will 
    internally estimate the S/N and relate it with a proper \textit{rejt} value. For a 
    detailed description of this new option, see Sect. \ref{AutoRejt}.
    
\vspace{0.2cm}  
\textbf{\textit{rejt} dependence on wavelength}: 
    The S/N can change significantly over the wavelength range for some spectra. With this in mind, 
    we have implemented a way to change the \textit{rejt} parameter for different wavelengths. 
    To use this option, the \textit{rejt} parameter has to be defined with the value $-2$. 
    ARES will then read a file with the name ``lambda\_rejt.opt', which should 
    be an ASCII file with two columns (wavelength and \textit{rejt} separated by spaces). 
    ARES will read the values in the file and then interpolate the \textit{rejt} value for each wavelength and will adjust 
    the continuum accordingly for each spectral region. 
    
\vspace{0.2cm}  
\textbf{EW error estimation}: One of the caveats in the original code was that it lacked an error 
    estimation or the flagging to identify problematic measurements. 
    This new feature will certainly be helpful, especially in spectral analyses where only very few lines are available. We have 
    implemented an efficient estimation of the error for the EWs. For more details on this new task see Sect. \ref{EWerror}.

\vspace{0.2cm}  
\textbf{Read ASCII spectral data}: Until now, ARES was only able to read specific FITS spectral data. 
    Since there are several spectra available in public archives that follow different formats (e.g., the ESPADONS or NARVAL archives), the new ARES code 
    can now read ASCII data. If a FITS file is not provided, ARES will try to read the file as ASCII data. For this to work, the data 
    should be available in two distinct columns (wavelength and flux separated by spaces). 

\vspace{0.2cm}  
\textbf{Cleaning any zero spectral flux}: One of the first tasks in ARES is to determine whether the input spectra have zero values in the flux. This 
    usually appears in spectral data from spectrographs that have a gap in the 
    spectral coverage. In many cases, the standard procedure for these cases is to assign zero values for these spectral regions. 
    Unfortunately, zero values in the flux can be problematic for the spectral normalization, and these regions should be avoided 
    to prevent the code from crashing. To overcome this problem, ARES scans the spectral data and replaces them for any zero value 
    by unity, which avoids unexpected crashes.
    
\vspace{0.2cm}  
\textbf{Parallelization of the code with OpenMp}: ARES is now implemented to use OpenMp. This is a code 
    parallelization library\footnote{\url{http://www.openmp.org}} that simplifies the interface with multicore machines. On 
    the ARES webpage we provide some benchmark examples that were used to compare and test the performance of the new version with 
    the older one. To summarize, these tests improve the performance from $\sim$ 60\% for K stars up to 
    $\sim$ 250\% for the F and G stars.

All these features are now implemented in ARES in a way that the code remains fully compatible with the previous version. The user 
is free to choose or discard any of the new features except for the errors, which will always appear in the new ARES output 
without any significant loss of performance; quite to the contrary, performance will improve if the code is used in a multicore machine.

\section{Automatic rejt}
\label{AutoRejt}

The identification of the continuum position is a very difficult and subjective task in these spectral analyses. The \textit{rejt} 
parameter plays a fundamental role in ARES for this purpose (see Fig. 2 and Sect. 3.1 of \citet[][]{Sousa-2007} for a complete 
description of this parameter). Because of that, this parameter still remains adjustable for the user. One of the 
advantages of automatic codes such as ARES is that when the parameter for the continuum is defined, the code will measure all lines in the spectra 
in a systematic way, in this way avoiding errors that are caused by the subjectivity that is common in the use of interactive routines.

\begin{figure*}[t]
  \centering
  \includegraphics[width=19cm]{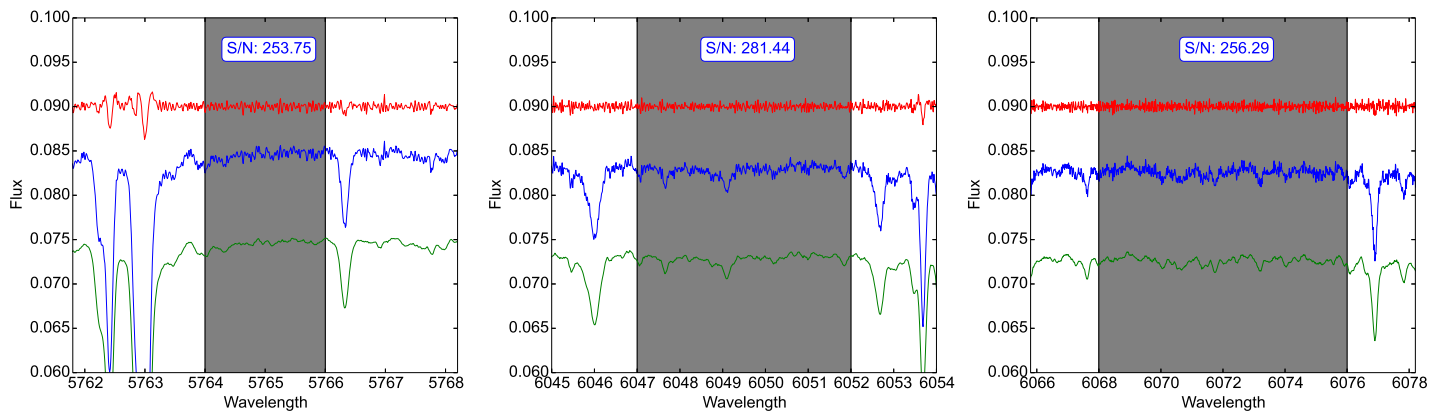}
  \caption{Evaluation of the noise in the three spectral regions recommended to automatically estimate the \textit{rejt} parameter (see text 
  for the details).}
  \label{FigAutoRejt}
\end{figure*}

To keep the EW measurements between different spectra homogeneous, the best option is to use an empirical relation between the spectral S/N 
and the \textit{rejt} parameter. This ensures the homogenization of the measurements for the analysis of several similar spectra
that were observed in the same spectroscopic configurations.

This solution was unsatisfactory, therefore we added the possibility of computing an automatic 
\textit{rejt} parameter. This is a difficult task, and some caution should be taken. This option can be accessed by 
defining the \textit{rejt} parameter in a different way. For example, 
setting \textit{rejt}$=$3;5764,5766,6047,6052,6068,6076 will tell ARES to evaluate the noise in three 
different spectral ranges, the first number (3) defines the number of ranges, and each next pair of numbers represents each wavelength region. 
These spectral regions should be selected carefully and avoid strong lines. The ideal case will be without any lines at all. The 
three spectral regions given in this example are regions that have no lines, or at most very week lines, for solar-type stars, and 
therefore we recommend them.

\subsection{Estimating the signal-to-noise ratio}

To obtain an automatically determined \textit{rejt} parameter, we start with estimating 
the S/N. Figure \ref{FigAutoRejt} is used here to explain the steps of estimating the S/N:

\begin{itemize}
 \item The first step is extracting the spectral region. The original spectrum is represented in blue for all panels, 
 and the region is marked by the gray area in the panels.
 
 \item The second step is to smooth the original spectra. This is achieved through a boxing car average. This allows 
 filtering the noise at higher frequencies (in wavelength) \citep[for more details see Sect. 3.3 of][]{Sousa-2007}. The value 
 for the smoothing in this procedure was hard-coded in ARES with a value of eight pixels.
 The smooth spectra are represented in green shifted down by a constant for clarity. The idea 
 is that these spectra do not represent any lower frequency signal, where the spectral lines are included.
 
 \item Next we subtract from the original spectrum this smoothed spectrum that is free from high-frequency noise and place it at the spectrum 
 level (using the average of the flux for each spectral region). This is represented with the red lines, again 
 slightly shifted by a constant for clarity. These data in red represent the high-frequency noise that was used to 
 estimate the S/N for each spectral region.
 
 \item The S/N computed for each spectral region is simply its mean flux $<F>$ divided by the standard deviation of the 
 subtracted spectra $\sigma_F$ (spectra in red): $S/N = \frac{<F>}{\sigma_F}$
  
 \item The final adopted S/N value is derived by taking the median of the three values derived for this example. The choice of 
 the median is more preferable than the average since it is not affected by the outlying points (e.g., a cosmic-ray hit).
\end{itemize}

Figure \ref{FigAutoRejt} shows the estimation of the S/N for the three spectral ranges given as example. Here we used a 
HARPS spectrum from the archive for the star \object{Alpha Cen A}. This spectrum was observed on 2005-04-23, 09:57:42 UT.
A S/N of $\sim$ 275 is indicated in the header of this spectrum. This value is compatible with the median of the 
values presented in the figure (256.29).

\subsection{Auto-\textit{rejt} and ARES local continuum determination}

The simple relation between \textit{rejt}, the S/N, and the standard deviation of the noise is 
$rejt = 1 - \frac{1}{S/N} = 1 - \sigma_F$, using $<F> = 1$ for the normalized flux at the continuum.

The auto-\textit{rejt} option should be used with caution since the spectral ranges selected may be contaminated with 
unexpected spectral features, such as cosmic rays or telluric lines. The S/N can also be directly
included in the \textit{rejt} parameter. In this case, the S/N is not estimated and directly uses this relation. This is 
an excellent option if the user has a prior knowledge on the S/N\ of the spectra.

\begin{figure}[!t]
  \centering
  \includegraphics[width=8cm]{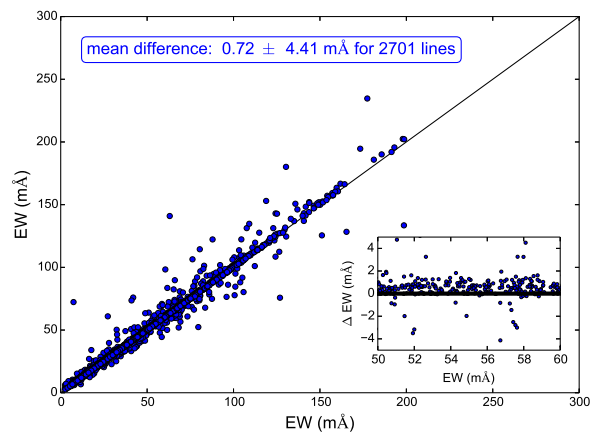}
  \caption{Comparison of the equivalent widths measured using the recommendation from \citet[][]{Sousa-2008} for the \textit{rejt} parameter and 
  using an automatic value.}
  \label{ewcomp2}
\end{figure}

To test the new feature, we present in Fig. \ref{ewcomp2} a comparison of EWs determined in two different ways. Here we used 
seven spectra with different S/N, ranging from $\sim$70 to $\sim$2000, measuring $\sim$ 380 lines for each of them. The goal is to 
show the differences betwen using the new auto-\textit{rejt} option with the option of using the recommended 
values from \citet[][]{Sousa-2008}. This comparison shows that there is a small mean offset of $\sim$ 0.7 m\AA\ with 
a still significant dispersion of $\sim$ 4.4 m\AA\. Most of the large spread is due to the spectra with lower S/N. The 
panel inside the figure shows a zoomed part of the comparison where it is clear that the high percentage ($\sim$ 96\%) of the 
lines are indeed close to the identity line.

We would like to recall that the importance of these automatic codes is their systematic performance. Regardless of the option 
selected in ARES or the value adopted for the \textit{rejt} parameter, the code will always behave systematically. This is very important 
for instance for a differential analysis, for which it is crucial that exactly the same method is used in the calibration and 
the spectral analysis. Another good example where the systematic measurements are very important is the use of line ratios to derive 
spectral indexes or even parameters such as the temperature \citep[][]{Kov-2003, Sousa-2010, Sousa-2012}.

\section{EW error estimation}
\label{EWerror}

There are many sources contributing to the EW error. These include the determination of the 
continuum position, the S/N of the data, blended lines, or any unexpected spectral feature, such as cosmic-ray hits.

An easy and fast solution was implemented in ARES to estimate the error on the equivalent width for each line. The error estimation is 
directly based on the fit that ARES performs on the spectral lines. Recalling this process, ARES fits the spectral lines with 
Gaussian profiles, using a non-linear square fit algorithm (gsl\_multifit\_fdfsolver\_lmder - defined in the Gnu Scientific library, GSL, as a 
robust and efficient version of the Levenberg-Marquardt 
algorithm\footnote{\url{http://www.gnu.org/software/gsl/manual/html_node/Minimization-Algorithms-using-Derivatives.html}}). To use this 
algorithm, the Gaussian profile and the respective parameter derivatives (the so-called Jacobian matrix) was implemented in the code. The Gaussian 
profile implemented in ARES is written as
\begin{equation}
 G(t) = \sum_i^n g_i(t) = \sum_i^n d_i \exp(-\Lambda_i (t - c_i))
,\end{equation}
where $d_i$ is the depth of the $i$th line, $c_i$ its center, and $\Lambda_i$ is related with the Gaussian width such 
as $\Lambda_i = 1 / (2 \sigma^2)$, where $\sigma$ is the Gaussian width (also known as the standard deviation). This specific representation of 
the Gaussian profile was chosen to closely follow the GSL example for the fitting of a single Gaussian with the non-linear square fit 
algorithm.
The difference to the previous ARES version is that the fit now considers an error for the spectral 
data. This is then used to estimate the errors on each parameter for the Gaussian fit. Note that although 
some reduced spectra in archives do have the error on the flux for each pixel, many do not. Because of this, and also to 
ensure consistency with the previous version of ARES, we still only require the input of the wavelength and the respective flux for each pixel. 
The error adopted for each pixel is taken as the same (every pixel will have the same weight), and its value strongly depends on the S/N, 
which is derived using exactly the same procedure as described in the previous section ($\sigma_F$). If
the \textit{rejt} parameter is defined by the user, $\sigma_F$ will be derived from its relation with the S/N.

With the uncertainty given for each of the fitted parameters, we can use it to estimate the error of a 
measurement. First, the equivalent width in ARES is computed directly from the area of the fitted Gaussian profile:
\begin{equation}
 EW_i = \int_{-\infty}^{\infty} g_i(t) dt = d_i \sigma_i \sqrt{2\pi} = d_i \sqrt{\frac{\pi}{\Lambda_i}}
,\end{equation}
where $i$ represents the Gaussian that fits the line to be measure.
Now we can use the uncertainties derived from the fit on the line depth ($\Delta d_i$) and on the line width ($\Delta \Lambda_i$), and 
from the error propagation equation we derive
\begin{equation}
 \Delta EW_i = EW_i \left[ \frac{\Delta d_i}{d_i} + \frac{1}{2} \frac{\Delta \Lambda_i}{\Lambda_i}\right] 
.\end{equation}
This error estimation includes the error on the continuum position (through the use of $\sigma_F$) and also part of the error 
introduced by blended lines that are responsible for the larger errors of the Gaussian profile parameters.

The main advantage of this new feature is that it is possible to immediately identify 
problematic EW measurements. This is very useful for spectral analysis, which relies on very 
few lines. For example, the analysis of elements such as Na, Mg, Al, Sc, and Mn are typically 
based only on two to four lines (per element) \citep[][]{Adibekyan-2012a}. The studies based on EW measurements for these elements 
using a limited line list will certainly benefit from evaluating the uncertaities for each line.

\section{Summary}

We presented a new version of the ARES code. This version includes new features. The most important are 
the automatic correction for radial velocity, the new option for automatically defining the continuum position, 
and the error estimation for the equivalent width measurements. All these features will certainly be 
useful for the users of ARES, especially for the analyses performed on large sets of spectral data.

The errors derived in ARES are good enough for a preliminary detection of poor measurements. This is a new 
capability, which can be used to pre-select the best-fit lines.  We still recommend 
the user to determine whether there are any outliers before using them for further analysis.

A more detailed description of the new features can be found on the ARES webpage. A proper 
understanding of these new features is fundamental for a proper judgment of the best option for each 
spectral analysis project. 

Finally, we would like to emphasize again that ARES v2 remains fully compatible with the previous version and will 
also work faster, thanks to the use of parallel processing for multicore processors.

\begin{acknowledgements}

This work is supported by the European Research Council/European 
Community under the FP7 through Starting Grant agreement
number 239953. N.C.S. was supported by FCT through the Investigador FCT 
contract reference IF/00169/2012 and POPH/FSE (EC) by FEDER funding 
through the program Programa Operacional de Factores de Competitividade - COMPETE. 
S.G.S, E.D.M, and V.A. acknowledge the
support from the Funda\c{c}\~ao para a Ci\^encia e Tecnologia, FCT (Portugal) and POPH/FSE (EC), in the
form of the fellowships SFRH/BPD/47611/2008, SFRH/BPD/76606/2011, and
SFRH/BPD/70574/2010. G.I. acknowledges financial support from the Spanish
Ministry project MINECO AYA2011-29060. 
\end{acknowledgements}

\bibliographystyle{bibtex/aa}
\bibliography{sousa_bibliography}

\end{document}